# Pressure effects on FeSe family superconductors

Yoshikazu Mizuguchi[a,b,c], Fumiaki Tomioka[a], Keita Deguchi[a,b,c], Shunsuke Tsuda[a,b], Takahide Yamaguchi[a,b] and Yoshihiko Takano[a,b,c]

[a]Superconducting Materials Center, National Institute for Materials Science, Tsukuba, 305-0047, Japan
[b]JST, Transformative Research-Project on Iron Pnictides, Tsukuba, 305-0047, Japan
[c]Graduate School of Pure and Applied Sciences, University of Tsukuba, Tsukuba, 305-0006, Japan

Corresponding author: Yoshihiko Takano
E-mail address: TAKANO.Yoshihiko@nims.go.jp
Address: 1-2-1 Sengen, Tsukuba 305-0821, Japan
Phone: +81-29-859-2842
Fax: +81-29-859-2601

Abstract
We investigated the pressure effects on the FeSe superconductor and the related compounds. Pressure dependence of superconducting transition temperature ($T_c$) for $FeSe_{0.8}S_{0.2}$ exhibits a dome-shaped behavior below 0.76 GPa. On the other hand, the $T_c$ of $FeSe_{0.25}Te_{0.75}$ linearly increases up to 0.99 GPa. Here we discuss the relation between the physical pressure effects and the chemical pressure effects on the FeSe system.

1. Introduction

Since the discovery of LaFeAsO$_{1-x}$F$_x$ superconductor [1], various types of iron-based superconductors, which have layered structure with FeP, FeAs, FeSe, FeTe layers, have been discovered [2-8]. FeSe system is one of the most attracting systems, because it has the simplest crystal structure among the iron-based superconductors. While a superconducting transition temperature $T_c$ of FeSe at ambient pressure is ~12 K, the $T_c$ increases to 37 K under high pressure [9-11]. $T_c$ also increases by partial substitutions of S or Te for the Se site [6,12]. These substitutions generate chemical pressures, because S, Se and Te have the same valence and the different ionic radius. Considering the huge positive pressure effect in $T_c$, compressing the lattice seems to be effective to enhance the $T_c$. However the increase of $T_c$ in FeSe$_{1-x}$S$_x$, which is chemically pressurized FeSe, is only 2 K, on the other hand, the increase of $T_c$ in FeSe$_{1-x}$Te$_x$, which is applied a negative chemical pressure, is the lager value of 5 K. In fact, the physical pressure effects and chemical pressure effects are not equivalent in FeSe system. Here we discuss pressure effects on FeSe, FeSe$_{1-x}$S$_x$ and FeSe$_{1-x}$Te$_x$ to clarify the factor which determines the $T_c$.

2. Materials and methods

Polycrystalline samples of FeSe, FeSe$_{0.8}$S$_{0.2}$ and FeSe$_{0.25}$Te$_{0.75}$ were synthesized using a solid-state reaction method as described in ref. 12. Temperature dependence of magnetization under pressure both in a zero-field cooling mode and a field cooling mode were measured using a SQUID magnetometer with an applied magnetic field of 10 Oe. The sample was placed in the center of a Cu-Be piston cylinder pressure cell which was filled with Fluorinert (FC70:FC77=1:1) as pressure transmitting medium. Small pieces of Pb or Sn was also placed in the pressure cell at a distance of 20 mm from the sample to determine the actual pressure by estimating its superconducting transition temperature. The magnetization data of a normal state was normalized to be zero.

3. Results and discussion

Figure 1 shows a temperature dependence of magnetization for FeSe under pressures, where the anomalies around 7 K are contributed by the superconducting transition of the Pb manometer. The $T_c$ increases monotonically with increasing pressure up to 0.67 GPa. The shielding volume faction increases with increasing pressure below 0.54 GPa and then saturates at 0.67 GPa. The pressure, where the shielding volume fraction becomes the maximum value, corresponds to the pressure at which the drop of the resistivity becomes the sharpest on the superconducting transition in pressurized FeSe. This behavior should be related to any change such as in the crystal structure, the electronic state or the magnetic fluctuation. In fact, the NMR measurements indicated a presence of a plateau in a pressure dependence of nuclear spin-lattice relaxation rate $1/T_1T$. The estimated $T_c$s are plotted in Fig. 2, which includes $T_c$s estimated from the resistivity measurements, as a function of pressure, and the values almost correspond to the $T_c^{mid}$ which is a mid point temperature in resistivity measurement.

Figure 3 shows a temperature dependence of magnetization for $FeSe_{0.8}S_{0.2}$, which shows the highest $T_c$ in $FeSe_{1-x}S_x$ as reported in ref. 12, under high pressure. Anomalies around 4 K are contributed by the superconducting transition of Sn manometer. The $T_c$ slightly increases up to 0.64 GPa, and then decreases under further pressure.

Figure 4 shows a temperature dependence of magnetization for $FeSe_{0.25}Te_{0.75}$. The sample is close to an end member FeTe, which is not superconducting and shows an antiferromagnetic ordering below 70 K. With applying pressure, the $T_c$ linearly increases in this pressure region, and reaches 14 K at 0.99 GPa.

The estimated $T_c$s of $FeSe_{0.8}S_{0.2}$ and $FeSe_{0.25}Te_{0.75}$ are plotted in Fig. 5 as a function of pressure. $FeSe_{0.8}S_{0.2}$ shows a dome-shaped behavior, which is similar to what is reported for the several iron-based superconductors. On the other hand, the $T_c$ of $FeSe_{0.25}Te_{0.75}$ monotonously increases still at 0.99 GPa and seems to increase under further pressure. To clarify the factor, which determines the $T_c$ in FeSe system, we discuss the relationship between the physical pressure effect and the chemical pressure effect in the following paragraph.

The S substitution for the Se site generates the positive chemical pressure, because S has the same valence with Se and the smaller ionic radius than that of Se. In $FeSe_{1-x}S_x$ system, the $T_c$ increased with S substitution up to 20 %, and then decreased with further substitution[12]. If $FeSe_{0.8}S_{0.2}$ is the compound, which is optimally pressurized by the S substitutions, it should not show an increase of $T_c$ under pressure. Contrary to this expectation, the $T_c$ of $FeSe_{0.8}S_{0.2}$ increases under pressure up to ~0.6 GPa. These results indicate that the structural changes, which are yielded by the S substitutions, are not fully equivalent with the structural changes which are yielded by applying pressure.

The Te substitution for the Se site generates the negative chemical pressure, because Te has the same valence with Se and the larger ionic radius than that of Se. Although the Te substitution expands the lattice, the $T_c$ increased with Te substitution and reached 14 K for $FeSe_{1-x}Te_x$ [6,12,13] in spite of the negative chemical pressure.

Furthermore the $T_c$ reaches above 20 K under high pressure in both $Fe_{1+\delta}Se_{0.57}Te_{0.43}$ [14] and $FeSe_{0.5}Te_{0.5}$ [15]. In this work, we observed the enhancement of $T_c$ under pressure also for $FeSe_{0.25}Te_{0.75}$. These large enhancements of $T_c$ under pressure in $FeSe_{1-x}Te_x$ imply that the structural change yielded by the Te substitution is disadvantageous for higher $T_c$. There should be another factor which enhances the $T_c$. One candidate to explain these features is the change in stability of magnetic ground states. Density functional calculation suggested that the higher stability of spin density wave ground state enhances the $T_c$ in $FeSe_{1-x}Te_x$, and the $T_c$ in the FeTe-based compounds will be higher than that of FeSe [16]. Superconductivity at higher $T_c$ will appear in FeSe family by optimizing these factors.

4. Conclusion

We investigated the pressure effects on FeSe and the related compounds. The $T_c$s of FeSe increased with applying pressure up to 0.67 GPa. The $T_c$ of $FeSe_{0.25}Te_{0.75}$ also showed the similar pressure dependence, and the $T_c$ will increase with applying further pressure. On the other hand, the pressure dependence of $T_c$ for $FeSe_{0.8}S_{0.2}$ exhibited the dome-shaped behavior. These results imply that the physical pressure effect and chemical pressure effect are not equivalent in FeSe system. More detailed study on the pressure effects for $FeSe_{1-x}S_x$ and $FeSe_{1-x}Te_x$, which are applied chemical pressures, should be addressed. It will provide us several important clues to elucidate the factors which determine the $T_c$ of iron-based superconductors.

Acknowledgements

This work was partly supported by a Grand-in-Aid for Scientific Research (KAKENHI).

References

[1] Y. Kamihara et al., J. Am. Chem. Soc. 130 (2008) 3296.

[2] Z. A. Ren et al., Europhys. Lett. 82 (2008) 57002

[3] M. Rotter et al., Phys. Rev. Lett. 101 (2008) 107006.

[4] X. C. Wang et al., Solid State Commun. 148 (2008) 538.

[5] F. C. Hsu et al., Nat. Acad. Sci. USA 105 (2008) 14262.

[6] M. H. Fang et al., Phys. Rev. B 78 (2008) 224503.

[7] Y. Mizuguchi et al., Appl. Phys. Lett. 94 (2009) 012503.

[8] H. Ogino et al., Supercond. Sci. Technol. 22 (2009) 075008.

[9] Y. Mizuguchi et al., Appl. Phys. Lett. 93 (2008) 152505.

[10] S. Margadonna et al., Phys. Rev. B 80 (2009) 064506.

[11] S. Medvedev et al., Nat. Mater. 8 (2009) 630.

[12] Y. Mizuguchi et al., J. Phys. Soc. Jpn. 78 (2009) 074712.

[13] B. C. Sales et al., Phys. Rev. B 79 (2009) 094521.

[14] N. C. Gresty et al., submitted to J. Am. Chem. Soc.

[15] K. Horigane et al., J. Phys. Soc. Jpn. 78 (2009) 063705.

[16] A. Subedi et al., Phys. Rev. B 78 (2008) 134514.

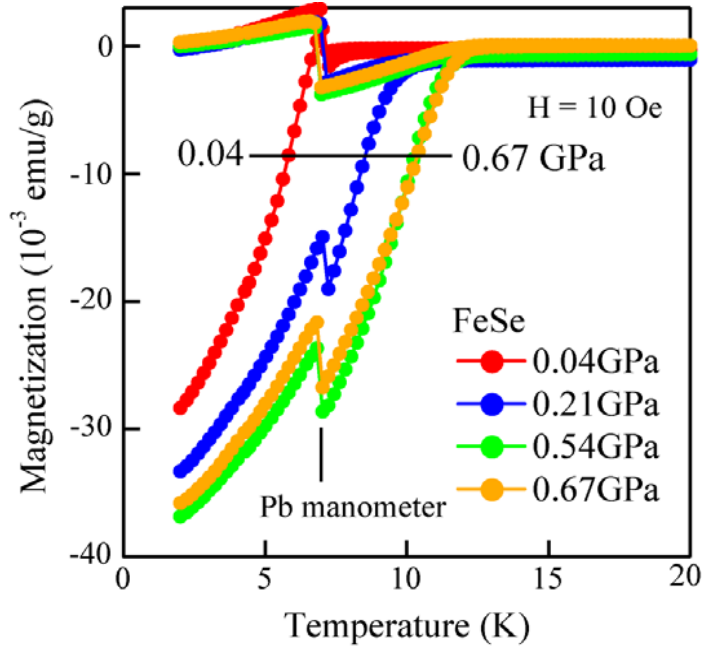

Fig. 1. Temperature dependence of magnetization under high pressures for FeSe. The anomalies around 7 K are attributed to the superconducting transitions of the manometer Pb.

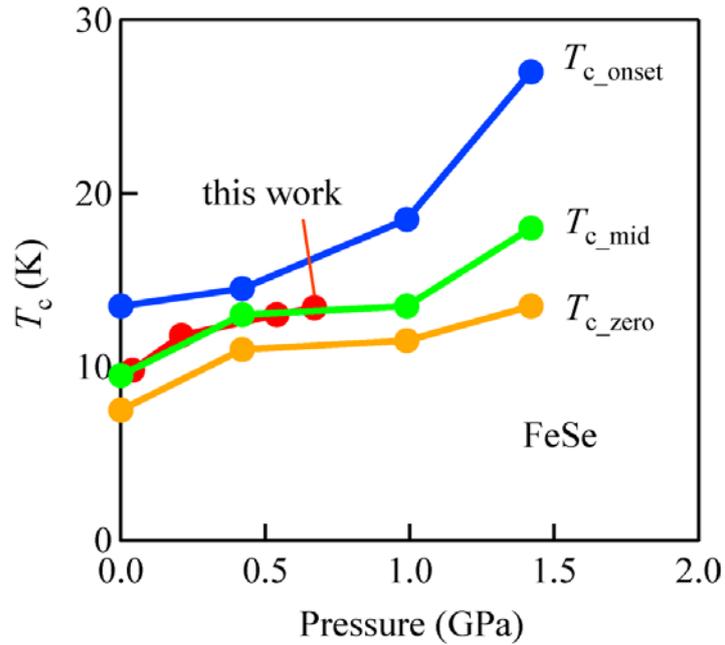

Fig. 2. Pressure dependence of $T_c$ for FeSe. The $T_c$s estimated from magnetization measurements are plotted with the $T_c$s estimated from resistivity measurements reported in ref. 9.

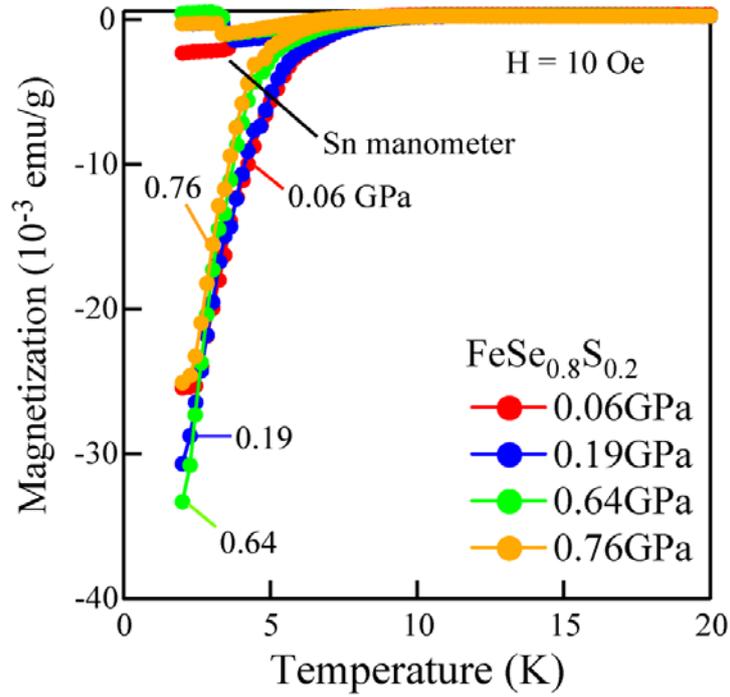

Fig. 3. Temperature dependence of magnetization under high pressures for FeSe$_{0.8}$S$_{0.2}$. The anomalies around 3 K are attributed to the superconducting transitions of the manometer Sn.

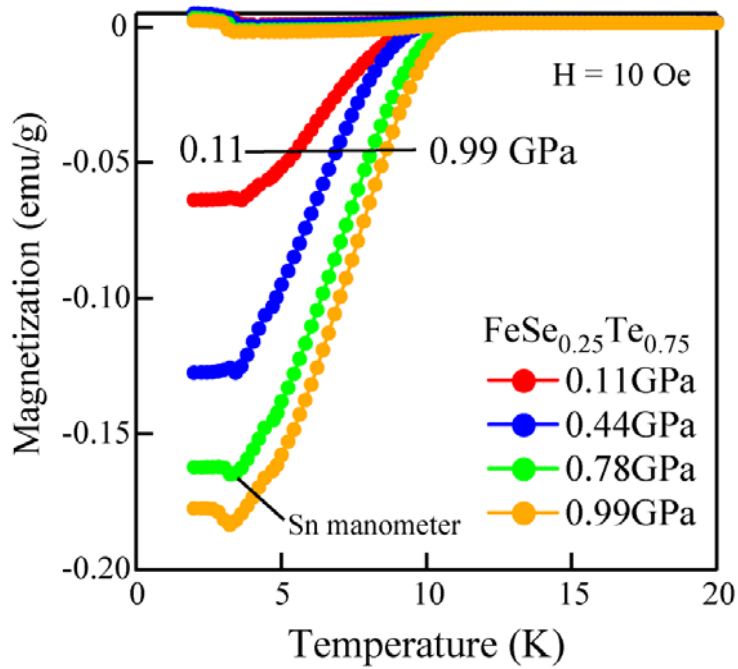

Fig. 4. Temperature dependence of magnetization under high pressures for FeSe$_{0.25}$Te$_{0.75}$. The anomalies around 3 K are attributed to the superconducting transitions of the manometer Sn.

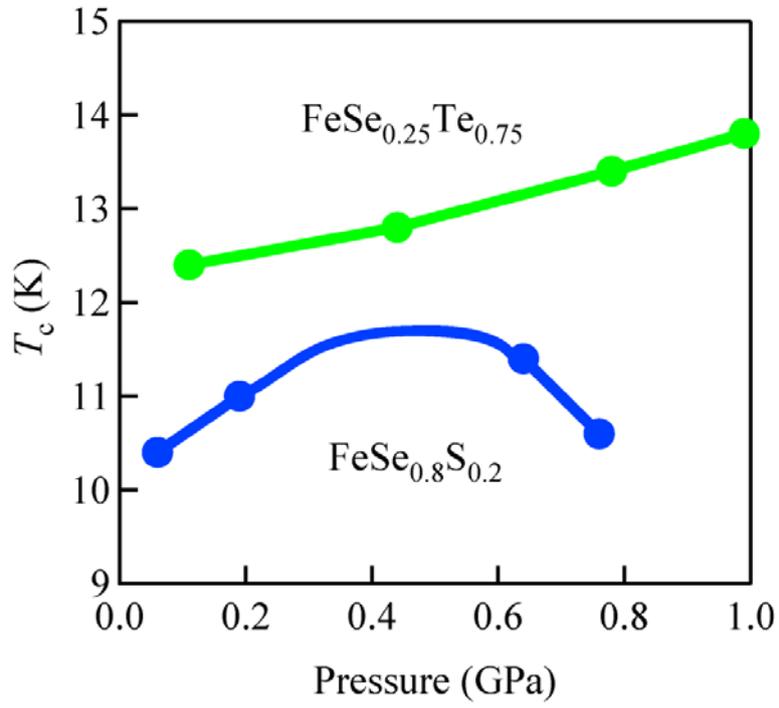

Fig. 5. Pressure dependence of $T_c$ for $FeSe_{0.8}S_{0.2}$ and $FeSe_{0.25}Te_{0.75}$. $FeSe_{0.8}S_{0.2}$ shows the dome-shaped behavior. For $FeSe_{0.25}Te_{0.75}$, the $T_c$ monotonously increases up to 0.99 GPa.